\documentclass[preprint,12pt]{elsarticle}

\usepackage{bm}
\usepackage{amsmath,amssymb,amsfonts,times}
\usepackage{url}

\usepackage[colorlinks=true, pdfstartview=FitV, linkcolor=red, citecolor=blue, urlcolor=blue]{hyperref}

\usepackage{graphicx}
\usepackage{epsfig}
\usepackage{slashed}
\usepackage{a4wide}
\usepackage{fancyhdr}
\usepackage{subfigure}

\usepackage{rotating}

\usepackage{pdflscape}

\allowdisplaybreaks[1]

\begin{document}

\newcommand{\UNIT}[1]{\mbox{$\,{\rm #1}$}}
\newcommand{\MeV}{\UNIT{MeV}}
\newcommand{\GeV}{\UNIT{GeV}}
\newcommand{\GeVc}{\UNIT{GeV/c}}
\newcommand{\TeV}{\UNIT{TeV}}
\newcommand{\AMeV}{\UNIT{AMeV}}
\newcommand{\AGeV}{\UNIT{AGeV}}
\newcommand{\ATeV}{\UNIT{ATeV}}
\newcommand{\fm}{\UNIT{fm}}
\newcommand{\mb}{\UNIT{mb}}
\newcommand{\nb}{\UNIT{nb}}
\newcommand{\fmc}{\UNIT{fm/c}}
\newcommand{\proz}{\UNIT{\%}}
\newcommand{\ds}{\displaystyle}
\newcommand{\zerovec}{\vec{0}\,}
\newcommand{\qvec}{\vec{q}\,}
\newcommand{\kvec}{\vec{k}\,}
\newcommand{\lvec}{\vec{l}\,}
\newcommand{\nvec}{\vec{n}\,}
\newcommand{\pvec}{\vec{p}\,}
\newcommand{\rvec}{\vec{r}\,}
\newcommand{\jvec}{\vec{j}\,}
\newcommand{\Gcapvec}{\vec{G}\,}
\newcommand{\Pcapvec}{\vec{P}\,}
\newcommand{\Picapvec}{\vec{\Pi}\,}
\newcommand{\Scapvec}{\vec{S}\,}
\newcommand{\Tcapvec}{\vec{T}\,}
\newcommand{\lambdavec}{\vec{\lambda}\,}
\newcommand{\sigmavec}{\vec{\sigma}\,}
\newcommand{\omegavec}{\vec{\omega}\,}
\newcommand{\sigmacapvec}{\vec{\Sigma}\,}
\newcommand{\thetacapvec}{\vec{\Theta}\,}
\newcommand{\tauvec}{\vec{\tau}\,}
\newcommand{\rhovec}{\vec{\rho}\,}
\newcommand{\deltavec}{\vec{\delta}\,}
\newcommand{\pivec}{\vec{\pi}\,}
\newcommand{\xivec}{\vec{\xi}\,}
\newcommand{\xvec}{\vec{x}\,}
\newcommand{\nrmsq}{\mbox{$\langle r^2\rangle_n$}}
\newcommand{\be}{\begin{equation}}
\newcommand{\ee}{\end{equation}}
\newcommand{\ba}{\begin{eqnarray}}
\newcommand{\ea}{\end{eqnarray}}
\newcommand{\etal}{\mbox{\it et al.}}
\newcommand{\prim}{\hspace{-1mm}'}
\def\bm#1{\mbox{\boldmath$#1$} }
\newcommand{\rp}{\textsc{r/p}-}
\newcommand{\derivl}{\stackrel{\leftarrow}{\partial}}
\newcommand{\derivr}{\stackrel{\rightarrow}{\partial}}

\newcommand{\evl}{e^{\frac{i\stackrel{\leftarrow}{\partial}_{\beta}v^{\beta}+m}{\Lambda}}}
\newcommand{\evr}{e^{\frac{-v^{\beta}i\stackrel{\rightarrow}{\partial}_{\beta}+m}{\Lambda}}}
\newcommand{\esl}{e^{\frac{i\stackrel{\leftarrow}{\partial}_{\beta}v^{\beta}+m}{\Lambda}}}
\newcommand{\esr}{e^{\frac{-v^{\beta}i\stackrel{\rightarrow}{\partial}_{\beta}+m}{\Lambda}}}

\newcommand{\err}{e^{\frac{-v^{\beta}i\stackrel{\rightarrow}{\partial}_{\beta}+m}{\Lambda}}}
\newcommand{\Gv}{\frac{g_{\omega}}{\Lambda}}
\newcommand{\Gr}{\frac{g_{\rho}}{\Lambda}}
\newcommand{\Gs}{\frac{g_{\sigma}}{\Lambda}}

\newcommand{\evnml}{e^{\frac{E}{\Lambda}}}
\newcommand{\esnml}{e^{\frac{E}{\Lambda}}}

\newcommand{\evnm}{e^{-\frac{E-m}{\Lambda}}}
\newcommand{\esnm}{e^{-\frac{E+m}{\Lambda}}}

\newcommand{\Gva}{\frac{g_{\omega}}{\Lambda}}
\newcommand{\Gsa}{\frac{g_{\sigma}}{\Lambda}}

\newcommand{\Gvb}{\frac{g_{\omega}}{ (\Lambda)^{2}}}
\newcommand{\Gsb}{\frac{g_{\sigma}}{ (\Lambda)^{2}}}

\newcommand{\fac}{\frac{\kappa}{(2\pi)^{3}}}

\newcommand{\nld}{{\cal D}}
\newcommand{\calo}{{\cal \Omega}}

\newcommand{\nldl}{\overleftarrow{{\cal D}}}
\newcommand{\nldr}{\overrightarrow{{\cal D}}}

\newcommand{\calol}{\overleftarrow{\varOmega}}
\newcommand{\calor}{\overrightarrow{\varOmega}}

\newcommand{\pspace}{\int\limits_{|\pvec|\leq p_{F_{i}}}\!\!\!\!\!\! d^{3}p}

\newcommand{\partialr}{\overrightarrow{\partial}}
\newcommand{\partiall}{\overleftarrow{\partial}}

\newcommand{\xil}{\overleftarrow{\xi}}
\newcommand{\xir}{\overrightarrow{\xi}}

\newcommand{\chil}{\overleftarrow{\chi}}
\newcommand{\chir}{\overrightarrow{\chi}}

\newcommand{\dleftp}{\nldl^{\prime}\big|_{\hat{\xi}=0} \xil^{\alpha}}

\newcommand{\drightp}{\nldr^{\prime}\bigg|_{\hat{\xi}=0}\xir^{\alpha}}

\newcommand{\dleftpp}{\nldl^{\prime\prime}\big|_{\hat{\xi}=0} \xil^{\alpha}\xil^{\beta}}

\newcommand{\drightpp}{\nldr^{\prime\prime}\bigg|_{\hat{\xi}=0}\xir^{\alpha}\xir^{\beta}}

%%% e-exp
\newcommand{\forma}{e^{-\zeta^{\alpha}i\partialr_{\alpha}+m/\Lambda}}
\newcommand{\formanm}{e^{-\frac{E-m}{\Lambda}}}
%%% p-monopole
\newcommand{\formb}{\frac{1}{1+\sum_{j=1}^{4}\left(\zeta_{j}^{\alpha} \, i\partialr_{\alpha}\right)^{2}}}
\newcommand{\formbnm}{\frac{\Lambda^2}{ \Lambda^2+\vec{p}^{\,2}}}
%%% p-dipole
\newcommand{\formc}{\left[\frac{1}{1+\sum_{j=1}^{3}\left(\zeta_{j}^{\alpha} \, i\partialr_{\alpha}\right)^{2}}\right]^{2}}
\newcommand{\formcnm}{\left[\frac{1}{ 1+\left(\frac{p}{\Lambda}\right)^{2}}\right]^{2}}

\newcommand{\NPA}{Nucl.~Phys.~}
\newcommand{\PR}{Phys.~Rep.~}   
\newcommand{\PL}{Phys.~Lett.~}
\newcommand{\PRC}{Phys.~Rev.~}
\newcommand{\PRL}{Phys.~Rev.~Lett.~}
\newcommand{\EPJ}{Eur.~Phys.~J.~}

\newcommand{\Bla}{\Big<}
\newcommand{\Bra}{\Big>}
%\maxdeadcycles=100000

\newcommand{\panda}{$\overline{\mbox P}$ANDA~}

%%%%%%%%%%%%%%%%%%%%%%%%%%%%%%%%%%%%%%%%%%%%%%%%%%%%%%%%%%%%%%%%%%
\begin{frontmatter}

\title{Multi-Strangeness Production in Hadron Induced Reactions}

\author{T.~Gaitanos$^{1}$, Ch.~Moustakidis$^{1}$, G.A.~Lalazissis$^{1}$, H.~Lenske$^{2,3}$}
\address{$^{1}$ Dept. of Physics, Aristotle University of Thessaloniki, 55124 Thessaloniki, Greece}
\address{$^{2}$ Institut f\"ur Theoretische Physik, Universit\"at Giessen,
             D-35392 Giessen, Germany}
\address{$^{3}$ GSI Helmholtzzentrum f\"ur Schwerionenforschung, D-64291 Darmstadt, Germany}
\address{email: tgaitano@auth.gr}

\begin{abstract}
We discuss in detail the formation and propagation of multi-strangeness particles in 
reactions induced by hadron beams relevant for the forthcoming experiments at 
FAIR. We focus the discussion on the production of the decuplet-particle 
$\Omega$ and study for the first time the production and propagation mechanism of this 
heavy hyperon inside hadronic environments. The transport calculations show the possibility 
of $\Omega$-production in the forthcoming \panda-experiment, which can be achieved with 
measurable probabilities using high-energy secondary $\Xi$-beams. We predict cross sections 
for $\Omega$-production. The theoretical results 
are important in understanding the hyperon-nucleon and, in particular, the 
hyperon-hyperon interactions also in the high-strangeness sector. We emphasize the importance 
of our studies for the research plans at FAIR.
\end{abstract}

\begin{keyword}
\panda, $\bar{p}$-induced reactions, $\Xi$-induced reactions, double-$\Lambda$ hypernuclei, 
$\Xi$-hypernuclei, $\Xi$N interactions, $\Omega$-baryon, $\Omega$-production.
%PACS numbers: 21.65.-f, 21.65.Mn, 25.40.Cm
\end{keyword}
\end{frontmatter}

\date{\today}

%%%%%%%%%%%%%%%%%%%%%%%%%%%%%%%%%%%%%%%%%%%%%%%%%%%%%%%%%%%%%%%%%%%%%%%%%%%%%%%
\section{\label{sec1}Introduction}
%%%%%%%%%%%%%%%%%%%%%%%%%%%%%%%%%%%%%%%%%%%%%%%%%%%%%%%%%%%%%%%%%%%%%%%%%%%%%%%

Hadronic reactions induced by heavy-ion and hadron beams build the central tool 
to look deeper inside the hadronic equation of state (EoS). Of particular 
interest is the strangeness sector of the EoS. 
Baryons with strangeness degree of freedom modify the nuclear EoS significantly 
at compressions beyond saturation~\cite{NSEoSHyp1,NSEoSHyp2,NSEoSHyp3,NSEoSHyp4}. 
Such effects show up already in 
ordinary matter (finite nuclei). Adding hyperons to a nucleus typically leads to 
a rearrangement of the whole system. Although hyperons are fermions, they do not 
underly the Pauli-exclusion principle with nucleons because strangeness makes 
them distinguishable. As a consequence one observes  increased binding energies 
and even a slight shrinking of hypernuclei, corresponding to a larger saturation 
density~\cite{Hash:2006}.

Hyperons are also important for nuclear astrophysics. It is well known that 
the presence of hyperons in the cores of neutron stars may play important role 
in determining both bulk properties of neutron stars as well as various  
dynamical processes~\cite{EoSHyp1,EoSHyp2,EoSHyp3}. In particular, hyperons 
can be formed in the interior of neutron stars when 
the in-medium nucleon chemical potential is large enough to make the conversion 
of a nucleon into a hyperon energetically favorable. 
Actually, they can appear at densities of about $2-3$ times the saturation density 
of nuclear matter. This conversion relieves the Fermi pressure exerted by the 
nucleons and makes the equation of state softer. It has been found that, the 
mentioned softness of the equation of state, leads to low values of maximum neutron 
star mass. This is in contradiction with a very recent accurate measurements of the 
masses, $M=1.97\pm 0.04 M_{\odot}$ (PSR J1614-2230 \cite{Demorest}) and 
$M=2.01 \pm 0.04 M_{\odot}$ (PSR J0348+0432 \cite{Antoniadis}). This is the so 
called hyperon puzzle where while the presence of hyperons at high densities is 
predicted by the nuclear theory is not compatible with measured neutron star masses.

There are, mainly, three different approaches  to study the hyperon formation in 
neutron star matter. The first one is based in the framework of the 
Brueckner-Hartree-Fock approach by using realistic nucleon-nucleon and hyperon-hyperon 
interactions~\cite{Bombaci-016,BHF2,BHF3}. The second method is based on Relativistic 
Mean Field Theory~\cite{RMFHyp1,RMFHyp2} 
and the third one on the construction of an effective equation of state
by employing Skyrme-type interactions~\cite{SkHyp1}. 
It has been suggested that the
hyperon-hyperon repulsion and hyperonic three-body interactions effects may
help to solve the  hyperon puzzle problem~\cite{Bombaci-016}. 
Another recent review article on this still debated issue can be found in 
Ref.~\cite{Chatterjee} (and further references therein).  

The hyperon-puzzle is one of the most recent issues concerning the study of the static 
and dynamic properties of the neutron stars~\cite{Chatterjee,blaschke16,QMC}. 
The solution of this problem may lead to 
a much better understanding of a complex phenomena in neutron star interior, such as 
the hyperon superfluidity and the hyperon bulk viscosity. All the mentioned effects 
are directly related with the neutron star cooling process, the glitches  and the 
radiation of gravitation waves~\cite{cooling,Glendenning-2000,Weber-99,latti16,latti15}.

Heavy-ion 
collisions at intermediate relativistic energies of several \GeV~per particle 
supply information on the in-medium hadronic properties over a broad range in baryon 
density~\cite{Reisdorf,Hermann}. Heavy-ion reactions at energies around the 
strangeness production threshold have been studied theoretically and experimentally 
in the past~\cite{Fuchs,Hartnack}. One of the most important achievements was the 
conclusion of 
a soft nuclear EoS at densities $\rho_{B}\simeq (2-3)\rho_{sat}$, as a result from 
collective flow~\cite{Reisdorf} and kaon 
production~\cite{Fuchs,forster,cassing1,cassing2} studies. 
For further reading concerning the high density equation of state of symmetric 
matter and the symmetry energy of compressed matter we refer to 
Refs.~\cite{dani,tsang,watt}. 
Supplementary 
information on the in-medium properties of kaons have been reported also in 
Ref.~\cite{Agakishiev}. Further investigations on strangeness and hypernuclear 
production in hadronic reactions have been recently started by several theoretical and 
experimental groups~\cite{tomasik,hyphi15,alicia15,botv15}.

While intermediate-energy heavy-ion reactions give essential details of the highly 
compressed matter, the high production thresholds of heavier hyperons hinder 
their production. One the other hand, the main task of flavor nuclear physics consists 
in the construction of the in-medium interaction between the octet and decuplet 
baryons $N,~\Lambda,~\Sigma,~\Xi$ and 
$\Omega$~\cite{Haidenbauer,gal,Gibson:1995an,HypFirst1,SchaffnerBielich:2008kb}. 
A possible way to overcome the high production thresholds without going beyond 
the nucleonic environment (quarks) is to study hadron-induced reactions. Of particular 
interest are antiproton-beams because of their high annihilation cross sections at 
intermediate energies~\cite{PDG}. Antinucleon-nucleon annihilation into multiple 
meson production and hyperonic resonances are the most important channels. Strangeness 
mesons (antikaons) and resonances can accumulate energy and strangeness degree through 
secondary scattering and produce by multi-step processes heavier hyperons. 

Up to now little is known about the hyperon-nucleon interaction. The 
uncertainty increases as the strangeness degree of freedom grows. 
Experimental information on the free hyperon-nucleon interactions is accessible 
in the S=-1 sector ($\Lambda N$ and $\Sigma N$ 
channels)~\cite{S1a,S1b,Valcarce:2005em,S1d,S1e}, but already for 
the S=-2 channels involving the cascade hyperon the situation is still 
very sparse~\cite{S2data}. Concerning the $S=-3$-processes with the 
$\Omega$-baryon no experimental data still exist. Consequently, from the 
theoretical side the parameters of the bare YN-interactions in the $S=-1$ 
channel ($\Lambda N,~\Sigma N$) are better under control than those parameters 
in the $S=-2$ sector ($\Xi N,~\Lambda\Lambda,~\Lambda\Sigma$). Indeed, 
various theoretical approaches with similar results for the $S=0$ channel 
($NN$-interaction) predict quite different results for the bare 
$\Xi N$-channels~\cite{S1d,S2data}. 

We have investigated in the past the formation and production mechanisms of 
hyperons~\cite{Larionov:2011fs}, fragments~\cite{Gaitanos:2007mm} 
and hyperfragments~\cite{Gaitanos:2013rxa,Gaitanos:2009at} in reactions 
induced by heavy-ions, protons and antiprotons. Recently the role of the 
multi-strangeness hyperon-nucleon interaction ($S=-2$) has been explored 
in detail~\cite{Gaitanos14}. It turned out that different hyperon-nucleon interaction 
models lead to observable effects and may constrain the high strangeness YN and 
YY interactions at \panda. In this work we extend our previous studies by 
considering the possibility of $\Omega$-production. The $\Omega$-baryon 
consists of three strange quarks preventing abundant $\Omega$-production in 
antiproton induced reactions. However, secondary strangeness exchange processes 
can increase the production of this heavy baryon. After an introduction to the 
theoretical aspects in Sec.~2 we discuss possible reaction mechanisms for 
$\Omega$-production in primary and secondary chance binary collisions (Sec.~3). 
In Sec.~4 we present the results for reactions involved in the 
\panda-experiment~\cite{panda1,panda2}.
That is, antiproton-induced reactions including the supplementary step with 
secondary $\Xi$-beams. We conclude the possibility of $\Omega$-formation in the 
reactions with the secondary beam at high energies above the $\Omega$-production 
threshold.

%%%%%%%%%%%%%%%%%%%%%%%%%%%%%%%%%%%%%%%%%%%%%%%%%%%%%%%%%%%%%%%%%%%%%%%%%%%%%%%%%%%
\section{\label{sec2} Hyperon-nucleon interactions}
%%%%%%%%%%%%%%%%%%%%%%%%%%%%%%%%%%%%%%%%%%%%%%%%%%%%%%%%%%%%%%%%%%%%%%%%%%%%%%%%%%%

In this Section we briefly outline the current status of the $S=-1$ and $S=-2$ 
in-medium octet-interactions before discussing the relevant part of the 
$\Omega$-baryon production. That is, the octet- and decuplet-interactions in 
the $S=-3$ sector.

\subsection{Hyperon-nucleon interaction in the $S=-1$ sector}

The hyperon-nucleon interaction is not yet fully understood for the entire baryon octet 
and decuplet. However, it is better fixed in the S=-1 sector from studies on 
single-$\Lambda$ 
hypernuclei~\cite{S1hyper1,S1hyper2,S1hyper3,S1hyper4,S1hyper5,S1hyper6,S1hyper7,
S1hyper8,S1hyper9,S1hyper10} 
and from $\Lambda$ (and $\Sigma$) production 
in reactions induced by hadrons and heavy-ions~\cite{Gaitanos:2013rxa,steinmeier,botv1,botv2}. 
The free hyperon-nucleon interaction in the S=-1 channels has been investigated in detail within 
one-boson-exchange (OBE) approaches~\cite{Nijmegen,Yama:2014,Juelich}. 
The rare available experimental data 
allowed to fix the various cross sections for a variety of elementary 
channels~\cite{Nijmegen,Fujiwara}. In-medium scattering has been also performed within the same 
OBE-scheme and in the spirit of relativistic mean-field (RMF) and density dependent 
hadronic (DDH) approaches~\cite{lenske}. 
The parameters are determined by simultaneous fits to NN- and YN-scattering 
observables in the S=-1 sector. While in the S=0 sector (NN-interactions) 
there are about 4300 scattering data of high quality available, in the S=-1 YN-part only 
38 scattering data are accessible. They still allow a reasonable determination of the 
$\Lambda$N and $\Sigma$N model parameters. 
Thus, despite of model differences all the theoretical approaches yield similar 
predictions for, e.g., potential depths and scattering cross sections for exclusive channels 
between nucleons and $\Lambda$-hyperons. Remaining uncertainties for 
S=-1 interactions are related to the unsatisfactory experimental data base. 

The basic task of determining the  full set of $YN$ and $YY$ hyperon interactions is 
still far from being under control to a satisfactory degree of accuracy. The progress 
made in the last decade or so for $S=-1$ systems is only part of the full picture 
since this involves mainly single-$\Lambda$ hypernuclei, supplementing the few data 
points from old $p\Lambda$ and $p\Sigma$ experiments, e.g. in 
\cite{Ale:1968,Sec:1968,Eis:1971}. The latter are essential input for the approaches 
developed over the years by several groups. While the Nijmegen \cite{Nijmegen,Yama:2014} 
and the Juelich group \cite{Juelich}, respectively, are using a baryon-meson 
approach, the Kyoto-Niigata group \cite{S1b} favors a quark-meson picture, 
finally ending also in meson-exchange interactions. 
However, none of the existing parameter sets is in any sense constrained with 
respect to interactions in the $|S|\geq 2$ channels. 

Furthermore we emphasize the theoretical developments based on lattice 
QCD simulations~\cite{lattice} and on the chiral 
effective field (EFT) theory~\cite{eft0}. Within the chiral EFT 
power counting prescriptions are employed to the $YN$-interactions~\cite{eft1,eft1b}. 
The chiral EFT approach has been 
recently applied to $Y(S=-1)N$-interactions in symmetric and asymmetric nuclear 
matter~\cite{eft1c}. Chiral EFT predictions for the real and imaginary parts of the 
in-medium optical $\Lambda$- and $\Sigma$-potentials are discussed. According to 
these studies, the in-medium $\Lambda$-potential is attractive while the corresponding 
$\Sigma$-potential is repulsive for single-particle momenta close to the 
Fermi-momentum. 

\subsection{Hyperon-nucleon interaction in the $S=-2$ sector}

Concerning the S=-2 sector only theoretical predictions exist so far in the 
literature. Among others, the chiral-unitary approach of Sasaki, Oset and 
Vacas~\cite{S1e} and the effective field theoretical models of the 
Bochum/J\"ulich groups~\cite{S1e} are representative examples. 
In this context we also mention the chiral effective field-theoretical 
approaches. They have been further extended to the $YN$-interactions 
in the $S=-2$ sector~\cite{eft3,eft4} with predictive power on $\Lambda\Lambda$ and 
$\Xi N$ cross sections. 
The Nijmegen~\cite{S1d} 
models are based on the well-known OBE picture to the NN-interaction, 
in which one then embeds the strangeness sector with the help of SU(3) symmetry. 
Fujiwara et al.~\cite{fuwiS2} have developed quark-cluster models for the 
baryon-baryon (BB) interactions in the S=0,-1,-2,-3,-4 sectors. 
Note that chiral EFT gives predictions on 
strangeness exchange cross sections such as $Y(S=-1)\Xi$ and $\Xi\Xi$ cross sections, 
see Refs.~\cite{eft3}. 
In the S=-2 sector involving the cascade-particles ($\Xi$), the uncertainties and, 
therefore, the discrepancies between the models are considerably larger~\cite{S2data}. 
We have investigated this task in a recent work~\cite{Gaitanos14} by choosing two particular model 
calculations for the hyperon-nucleon in-medium interaction in the S=-2 sector. For instance, 
the one-boson-exchange calculations of the Nijmegen group in the extended soft-core 
version ESC04 of 2004~\cite{S1d} and the quark-cluster approach~\cite{fuwiS2} give very 
similar predictions for the nucleon-nucleon interaction, but differ considerably between 
each other in the $\Xi N$-interactions. As discussed in detail in Ref.~\cite{Gaitanos14},
observable 
signals arise in reactions induced by $\Xi$-beams on intermediate mass target nuclei. These 
signals show up in the amount of bound cascade hyperons inside the target and, in particular, 
in the production of double-strangeness hypernuclei. 

\subsection{Hyperon-nucleon interaction in the $S=-3$ sector}

In the higher strangeness sector $S=-3$ the situation is still not understood. For reasons 
which will become clear below, we distinguish between primary and secondary production 
processes of the $\Omega$-baryon. For the primary binary processes we focus the discussion to 
antiproton-proton collisions. Various possibilities exist for the secondary $\Omega$-baryon 
production, which will be discussed below.

%%%%%%%%%%%%%%%%%%
\begin{figure}[t]
\vspace{-2cm}
\begin{center}
\includegraphics[clip=true,width=0.6\columnwidth,angle=-90.]{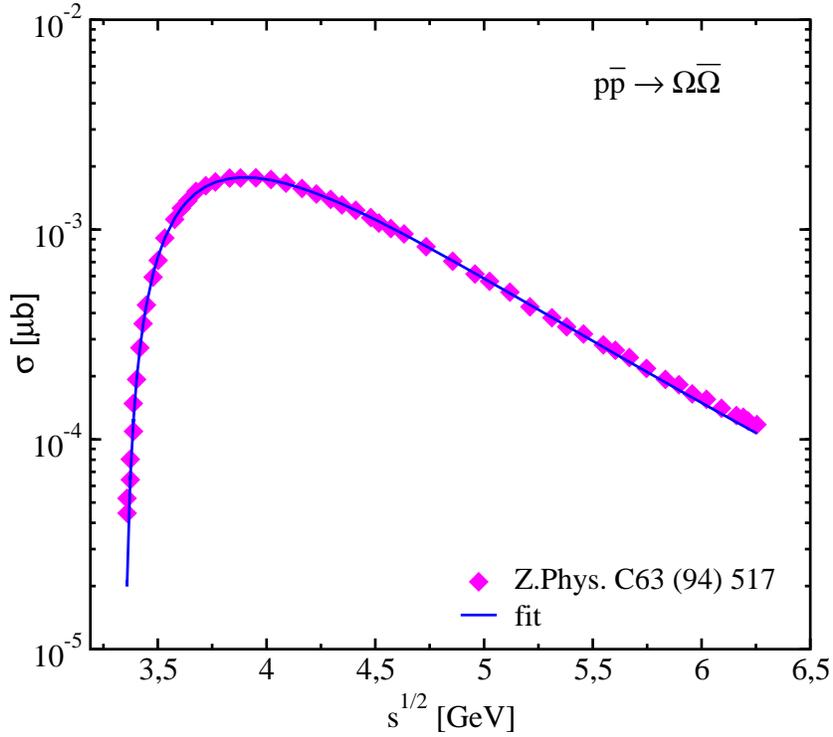}
%\vspace{-1.0cm}
\caption{\label{fig1}
Antiproton-proton cross section for $\Omega$-baryon production. The symbols refer to 
theoretical estimations from Ref.~\cite{kaidalov} and the solid curve is a
parametrization used in the transport calculations.
\vspace{-1.0cm}
}
\end{center}
\end{figure}
%%%%%%%%%%%%%%%%%%
For antiproton-proton annihilation into $\Omega$ particles one theoretical work exists in 
the literature only~\cite{kaidalov}, which is based on reggeon-like calculations for rather low 
energies of interest. The estimations of these calculations are shown in Fig.~\ref{fig1} in 
terms of the total production cross section (symbols). The solid curve is a polynomial 
parametrization to the theoretical results, which will be used in the transport descriptions. 
At first, one can see that the $\Omega$-production cross section is very low. Above the 
production threshold of $\sqrt{s}\geq 3.344$ \GeV~(corresponding to a kinetic energy of 
$E_{lab}=4$ \GeV~or to a beam momentum of $p_{lab}=4.9$ \GeV~in the laboratory frame) 
the cross section grows up to a maximum of several \nb~only before starting to decrease 
again. Note that these scales are approximately one (two) order of magnitude less with respect 
to the antiproton-proton production cross section of the cascade ($\Lambda$) particles, 
as shown and discussed in Ref.~\cite{kaidalov}.

One can naively explain the low production probability of the $\Omega$ particles, without 
going into theoretical details. In contrary to the $\Lambda$ and also the cascade 
hyperons, the decuplet $\Omega$-particle consists uniquely of three strange 
quarks. These have to be generated in intermediate flavor exchange processes during the 
antiproton-proton annihilation. Also the heavy $\Omega$-mass further reduces the phase-space 
with respect to the corresponding $\Lambda$ and $\Xi$ phase-space factors. Thus, the 
$\Omega$-production cross section becomes significantly lower. 

%%%%%%%%%%%%%%%%%%
\begin{figure}[t]
\vspace{-2cm}
\begin{center}
\includegraphics[clip=true,width=0.6\columnwidth,angle=-90.]{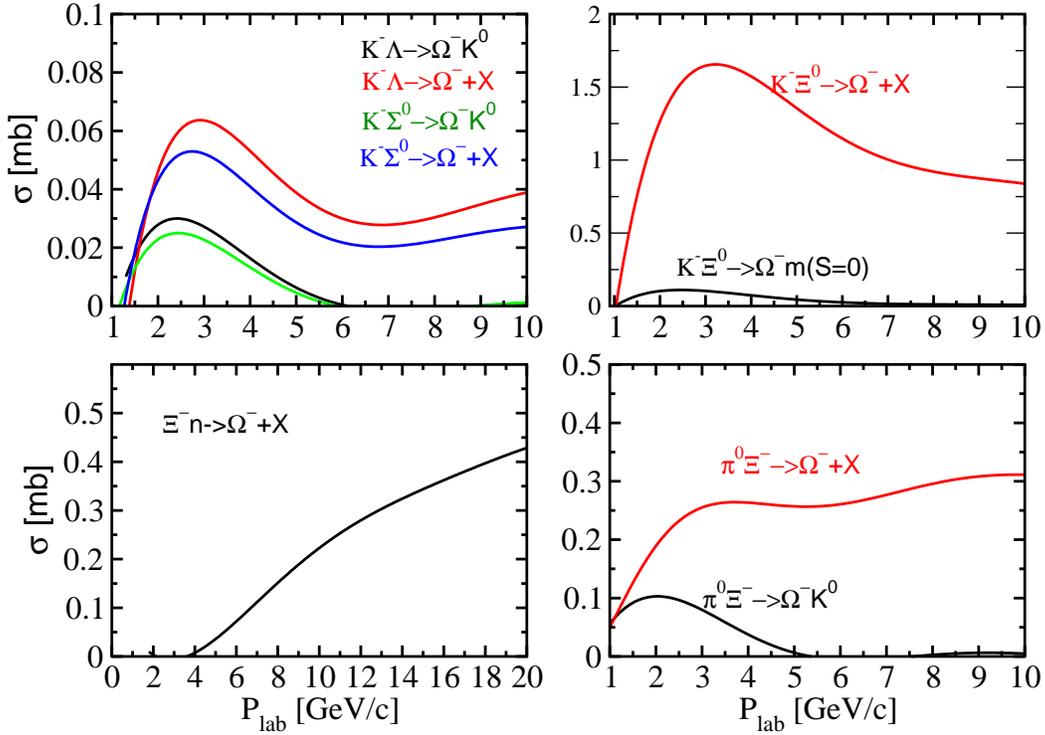}
%\vspace{-1.0cm}
\caption{\label{fig2}
Total cross sections for various channels (as indicated) for $\Omega$-baryon production.
\vspace{-1.0cm}
}
\end{center}
\end{figure}
%%%%%%%%%%%%%%%%%%
Consequently, the question arises whether the production of heavy hyperons with 
multi-strangeness degree of freedom is possible in the energy domain planned at FAIR. In this 
context one should remind, that the \panda-experiment consists of a two-step process: 
Antiproton-beams on a nuclear target as the origin of cascade particles ($\Xi^{-}$), which 
will serve as a $\Xi^{-}$-beam for collisions with a secondary target. Thus, secondary 
re-scattering may contribute to $\Omega$-production as well. Fig.~\ref{fig2} shows the 
cross sections for some particular secondary processes. Obviously we select such processes 
with the highest strangeness degree of freedom in the initial channel. That is, re-scattering 
between antikaons and other hyperons with $S=-1$ (panels on top in Fig.~\ref{fig2}) or 
between the cascade particles with other non-strange baryons or mesons (panels on the 
bottom in Fig.~\ref{fig2}). 
In principle, these cross sections can be evaluated within chiral EFT approaches. 
However, in view of the limited energy range accessible in the EFT approach and the 
persisting sizeable bands of uncertainty, in the present study we prefer to use 
phenomenological cross sections extracted directly from elementary events by means of the 
PYTHIA program~\cite{pythia}. 
At first, 
the particle energies should lie above the corresponding $\Omega$-production thresholds. 
With the $\Omega$-mass of $1.672$ \GeV~this results to a threshold of 
$\sqrt{s}\ge 3.107$ \GeV~for the $\Xi N\to \Omega K N$-channel 
(corresponding to $E_{lab}\ge 3.75$ \GeV~kinetic energy or to $p_{lab}\ge 3.5$ \GeV~beam 
momentum). According to these calculations high energy $\Xi$-beams are necessary to 
enforce the formation of $\Omega$-baryons from second chance binary collisions, as 
shown in the bottom-left panel of Fig.~\ref{fig2}. Note that the order of magnitude of 
these cross sections (\mb-region) is comparable with the typical strangeness 
production cross sections ($NN\to NYK$ and $\pi N\to YK$ with $Y=\Lambda,\Sigma$ and 
$K$ for the hyperons and kaons, respectively). Furthermore the $\Omega$-production cross 
sections for primary 
$p\bar{p}\to \Omega\bar{\Omega}$-channels is several orders of magnitude less with respect 
to the re-scattering cross sections. Obviously, high energy $\Xi$-beams with rather 
heavy-mass nuclear targets might be necessary to produce $\Omega$-particles with 
high probability. This is the topic of the next section.

%%%%%%%%%%%%%%%%%%%%%%%%%%%%%%%%%%%%%%%%%%%%%%%%%%%%%%%%%%%%%%%%%%%%%%%%%%%%%%%%%%%
\section{\label{sec3} Hadron-induced reactions}
%%%%%%%%%%%%%%%%%%%%%%%%%%%%%%%%%%%%%%%%%%%%%%%%%%%%%%%%%%%%%%%%%%%%%%%%%%%%%%%%%%%

For the theoretical realization of hadron-induced reactions we use the widely 
established relativistic Boltzmann-Uheling-Uhlenbeck (BUU) transport 
approach~\cite{Botermans:1990qi}. The kinetic equations are performed 
numerically within the Giessen-BUU (GiBUU) transport model~\cite{Buss:2011mx}. 
The GiBUU equation is given by
%%%%%%%%%%%%%%%
\begin{equation}
\left[
k^{*\mu} \partial_{\mu}^{x} + \left( k^{*}_{\nu} F^{\mu\nu}
+ m^{*} \partial_{x}^{\mu} m^{*}  \right)
\partial_{\mu}^{k^{*}}
\right] f(x,k^{*}) = {\cal I}_{coll}
%\frac{1}{2(2\pi)^9} \nonumber\\
%& & \times \int \frac{d^3 k_{2}}{E^{*}_{{\bf k}_{2}}}
%             \frac{d^3 k_{3}}{E^{*}_{{\bf k}_{3}}}
%             \frac{d^3 k_{4}}{E^{*}_{{\bf k}_{4}}} W(kk_2|k_3 k_4)
% \left[ f_3 f_4 \tilde{f}\tilde{f}_2 -f f_2 \tilde{f}_3\tilde{f}_4
%\right] 
\,.
\label{rbuu}
\end{equation}
%%%%%%%%%%%%%%%
It describes the dynamical evolution of the $1$-body phase-space 
distribution function $f(x,k^{*})$ for the various hadrons under the influence of a 
hadronic mean-field (l.h.s. of Eq.~(\ref{rbuu})) and binary collisions 
(r.h.s. of Eq.~(\ref{rbuu})). For the mean-field we adopt the relativistic mean-field 
approximation of Quantum-Hadro-Dynamics~\cite{qhd}. The hadronic potential shows up 
in the transport equation through the kinetic $4$-momenta $k^{*\mu}=k^{\mu}-\Sigma^{\mu}$ 
and effective (Dirac) masses $m^{*}=M-\Sigma_{s}$. The in-medium self-energies, 
$\Sigma^{\mu} = g_{\omega}\omega^{\mu} + \tau_{3}g_{\rho}\rho_{3}^{\mu}$ and 
$\Sigma_{s} = g_{\sigma}\sigma$, describe the in-medium interaction of nucleons 
($\tau_{3}=\pm 1$ for protons and neutrons, respectively). 
The isoscalar, scalar $\sigma$, the isoscalar, vector $\omega^{\mu}$ and the 
third isospin-component of the isovector, vector meson field $\rho_{3}^{\mu}$ 
are extracted from the standard Lagrangian equations of motion~\cite{qhd}. 
For the model parameters (obvious meson-nucleon couplings) we use the 
$NL3$-parametrization, which includes non-linear self-interactions of the 
$\sigma$ field ~\cite{lala}. 
The meson-hyperon couplings at the mean-field level are obtained 
from the nucleonic sector using simple quark-counting arguments. 
The in-medium nucleon-, $\Lambda$-, $\Sigma$- and $\Xi$-potentials are 
$U_{N}=-46$, $U_{\Lambda}=-38$, $U_{\Sigma}=-39$ and $U_{\Xi}=-22$ (in units 
of \MeV), respectively, at saturation density and zero kinetic 
energy~\cite{Larionov:2011fs}. For the $\Omega$-mean field one could use again 
the quark-counting argument. However, as we will see below, directly produced 
heavy baryons like the $\Omega$ escape immediately from the nuclear environment, 
thus inhibiting the formation of bound states. Important for captured $\Omega$-particles 
inside nuclear matter will be secondary scattering processes.
For the in-medium potential of the $\Omega$-baryon we use here the same potential 
as that for the nucleons for simplicity. The collision 
term includes all necessary binary processes for (anti)baryon-(anti)baryon, 
meson-baryon and meson-meson scattering and annihilation~\cite{Buss:2011mx}. 
For more details of the corresponding 
mean-field and cross section parameters we refer to
Refs.~\cite{Larionov:2011fs,S1a,S1e,landolt,yamariken}. 
Relevant for the present work is the implementation of the new parametrizations for 
the $N\bar{N}\to \Omega\bar{\Omega}$-scattering, as discussed in the previous section. 
Having the cross sections for all exclusive elementary channels of interest, the 
collision integral of the transport equation is then modelled within conventional 
Monte Carlo methods. 

%%%%%%%%%%%%%%%%%%
\begin{figure}[t]
\vspace{-2cm}
\begin{center}
\includegraphics[clip=true,width=0.6\columnwidth,angle=-90.]{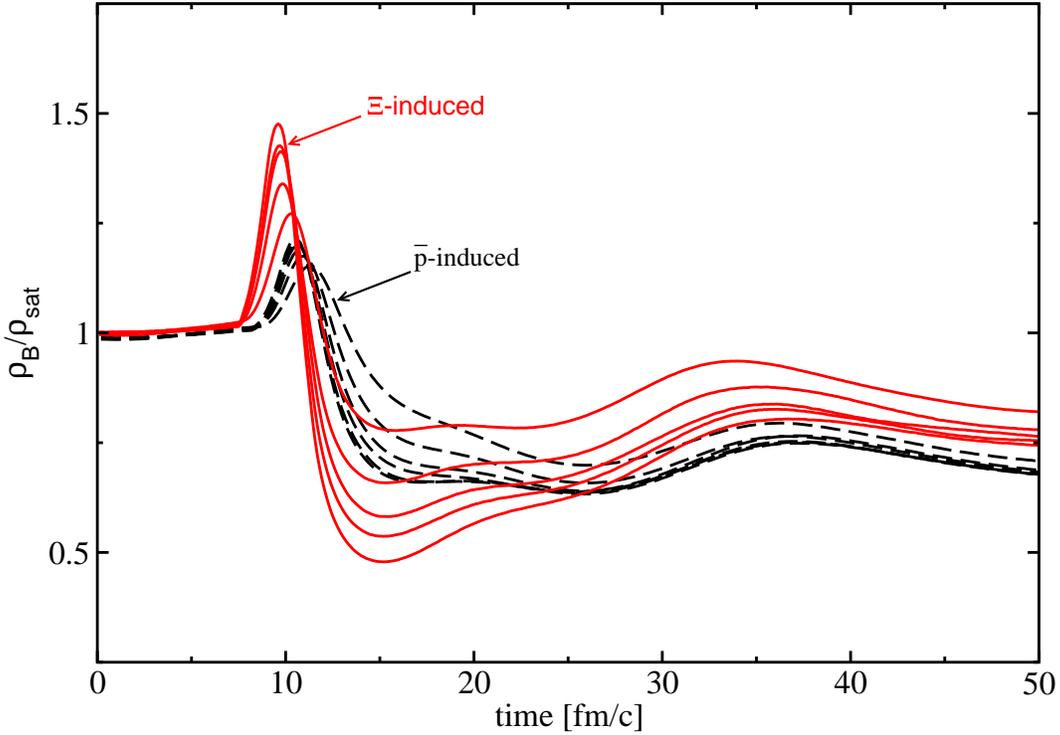}
%\vspace{-1.0cm}
\caption{\label{fig3}
Baryon density $\rho_{B}$ (in units of the saturation density $\rho_{sat}$) at the origin 
of the target nucleus as function of time for $\bar{p}+{}^{93}$Nb and $\Xi^{-}+{}^{64}$Cu 
reactions, as indicated.
\vspace{-1.0cm}
}
\end{center}
\end{figure}
%%%%%%%%%%%%%%%%%%
We have performed transport calculations for antiproton-induced reactions including 
the second-step process of $\Xi^{-}$-collisions on a secondary nuclear target. 
We have used two different target nuclei, ${}^{93}Nb$ and ${}^{64}Cu$ for the 
$\bar{p}$- and $\Xi$-induced reactions, respectively. For the antiproton-nucleus 
reactions a heavier target 
is used to increase the rare $\Omega$-production via secondary scattering, while 
in the $\Xi$-induced reactions a lighter target is sufficient for the same purpose. 
In 
extension to our previous work high energy secondary cascade-beam are considered here. 
Before analyzing the production of particles of interest, we discuss the density regions 
tested in such reactions. This is shown in Fig.~\ref{fig3} in terms of the temporal 
evolution of the baryon density for antiproton- and $\Xi$-induced reactions at 
different beam energies. The density is calculated at the origin of the corresponding 
nuclear target. Note 
that the compressions of the matter is less pronounced in $\bar{p}$-induced reactions with 
respect to the $\Xi$-nucleus collisions. This result seems surprising, since the in-medium 
antibaryon is very attractive in RMF, as discussed in Ref.~\cite{NLD1}. 
Also note that a heavier nucleus is used in the $\bar{p}$-induced reactions. 
However, the imaginary 
part of the in-medium antiproton optical potential is rather strong~\cite{NLD1,friedman}. 
It causes immediate annihilation already at the nuclear surface before deeper penetration 
of the antiproton. Consequently, the density averaged over the events does not differ much 
from the saturation value. Furthermore, 
apart the trivial time shifts between the various beam energies, both types of reactions 
show a similar 
behavior of the density evolution. During the hadron ($\bar{p},\Xi$) penetration into 
the nucleus binary processes cause a moderate compression at the center of the target. In the 
subsequent de-excitation stage the central baryon density decreases due to particle emission. 
In such reactions densities close to saturation $0.5\rho_{sat} \le \rho \le 1.5\rho_{sat}$ are
achieved and, thus, one can probe the in-medium interactions in this density range. 

%%%%%%%%%%%%%%%%%%
\begin{figure}[t]
\vspace{-2cm}
\begin{center}
\includegraphics[clip=true,width=0.6\columnwidth,angle=-90.]{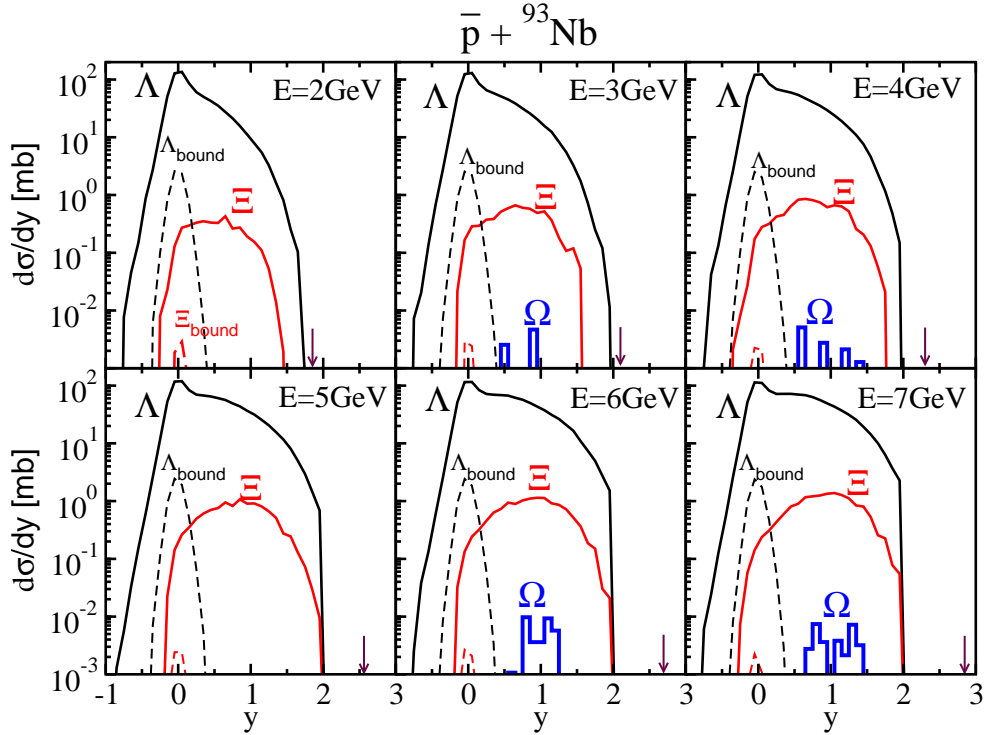}
%\vspace{-1.0cm}
\caption{\label{fig4}
Rapidity distributions for various hyperons (as indicated) produced in $\bar{p}+^{93}Nb$ 
reactions at antiproton-beam energies as shown in the panels. For the $\Lambda$ and $\Xi$ 
hyperons also the corresponding yields of them are displayed (dashed curves), which are 
bound inside the nucleus. The vertical arrows at each panel indicate the rapidity value 
of the corresponding beam-energy. 
\vspace{-1.0cm}
}
\end{center}
\end{figure}
%%%%%%%%%%%%%%%%%%
Fig.~\ref{fig4} shows the results of the transport calculations in terms of rapidity spectra 
of various produced hyperons for $\bar{p}$-induced reactions at different beam energies. 
We observe an abundant production of the lightest hyperon ($\Lambda$) with a broad spectral 
distribution 
in longitudinal momentum followed by the production of the cascade particles. The 
big differences in their production yields arise from the smaller annihilation cross section 
into cascade particles relative to that into the $\Lambda\bar{\Lambda}$-channel. Note that 
here secondary re-scattering is mainly responsible for the breadth of the rapidity 
spectra. It causes the capture of these hyperons inside the target nucleus ($\Lambda_{bound}$ 
and $\Xi_{bound}$ rapidity yields) with the subsequential formation of $\Lambda$-hypernuclei. 
We do not go into further details concerning the formation of hypernuclei. This task has been 
studied in the past in detail in previous works~\cite{Gaitanos14}. 

We focus now on the formation of the heavy $\Omega$-hyperon. It can be seen in Fig.~\ref{fig4} 
that the $\Omega$-production is a very rare process in antiproton-induced reactions at 
beam energies just close to the $\Omega$-production threshold of $\sqrt{s}\simeq 3.344$ \GeV. 
We emphasize that we have analyzed around $4$ millions of transport-theoretical events for 
each incident energy. 
The main reason for the low production yields of the $\Omega$-baryon is the extremely low 
annihilation cross section of several \nb~only. This value is far below the annihilation 
cross section of other exclusive processes. The major contribution to the annihilation 
cross section comes from multiple meson production~\cite{golu}. It is important to note that 
the origin of the produced $\Omega$-particles isn't $p\bar{p}$-annihilation, but other 
secondary processes involving re-scattering between antikaons, antikaonic resonances with 
hyperonic resonances ($Y^{\star}(S=-1)$). For instance, for the reaction $\bar{p}+{}^{93}Nb$ 
at $4$ \GeV~incident energy 
these secondary scattering processes contribute with a cross section of $1.148$ \nb~to 
the total $\Omega$-production yield with a cross section of $\sigma_{\Omega}=1.15$ \nb. 
The cascade particles and their resonances, 
which carries already $S=-2$ and thus would preferably contribute to the $\Omega$-formation, 
mainly escape the target nucleus. For this reason they do not contribute here. 

%%%%%%%%%%%%%%%%%%
\begin{figure}[t]
\vspace{-2cm}
\begin{center}
\includegraphics[clip=true,width=1.1\columnwidth,angle=0.0]{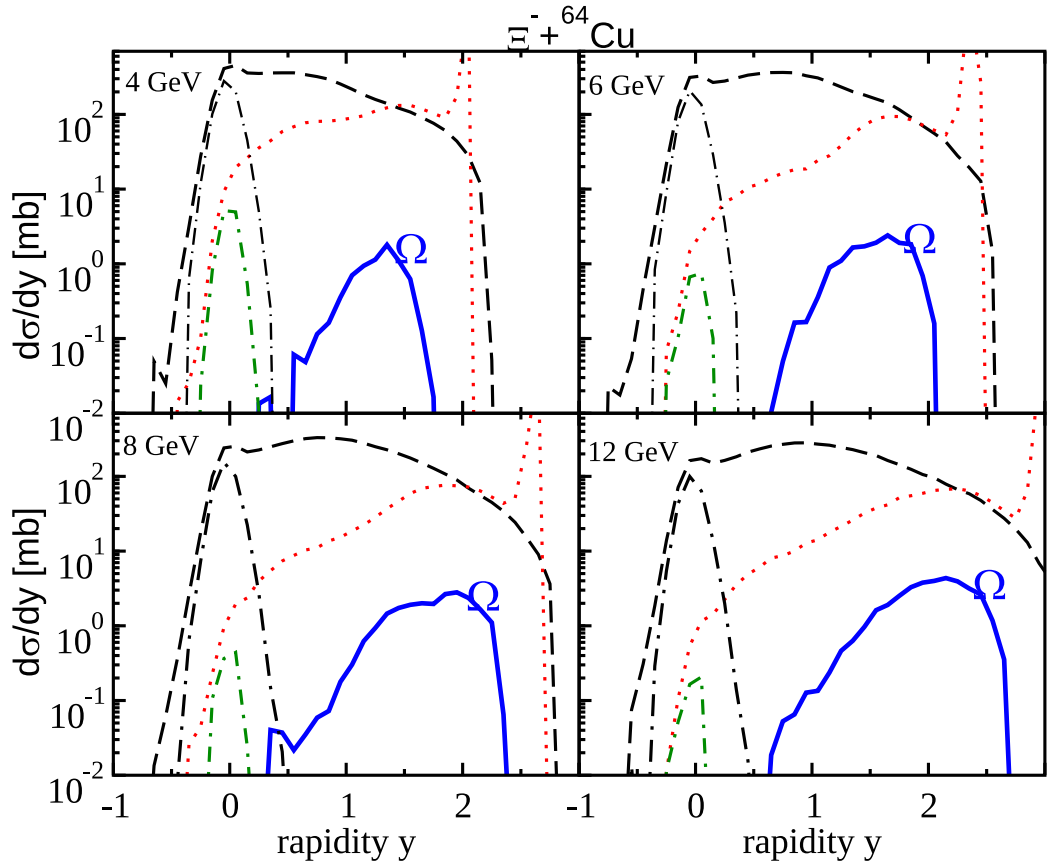}
\vspace{-1.0cm}
\caption{\label{fig5}
Rapidity distributions of different hyperons for $\Xi$-induced reactions on 
${}^{64}Cu$-targets at four beam-energies, as indicated in each panel. 
Dotted curve: $\Xi$-, dashed curve: $\Lambda$-, 
dott-dashed curve: bound $\Lambda$-, dot-dot-dashed curve: bound $\Xi$-, 
thick solid curve: $\Omega$-rapidity spectra. 
\vspace{-1.0cm}
}
\end{center}
\end{figure}
%%%%%%%%%%%%%%%%%%
The realization of a second target using the produced 
cascade particles as a secondary beam is important. At first, for the copious 
production of multi-strangeness hyperons and multi-strangeness hypernuclei, 
as proposed by the \panda-experiment~\cite{panda1,panda2}. 
According to the \panda-proposal low-energy $\Xi$-beams will be used for the production 
of $\Lambda\Lambda$-hypernuclei. First theoretical predictions on such exotic hypermatter 
in low-energy $\Xi$-induced reactions have been indeed reported in Ref.~\cite{Gaitanos14}. Not 
only $\Lambda\Lambda$-hypernuclei, but also the direct formation of $\Xi$-hypermatter is 
accessible depending on the cascade-nucleon interaction~\cite{Gaitanos14}. 

We show now that the same experiment can be used to explore the formation of $\Omega$-baryons. 
This decuplet-hyperon is heavy causing high production thresholds. As discussed above, secondary 
re-scattering including intermediate production of high-mass hyperonic resonances can be a 
more favorable 
possibility for $\Omega$-production. Due to the high strangeness value of this baryon the entrance 
channel should have as high as possible strangeness degree. Thus, the \panda-experiment with the 
secondary $\Xi$-beam can be a good candidate for our purpose. This is shown in Fig.~\ref{fig5} 
in terms of the rapidity spectra, but now for $\Xi$-induced reactions at higher incident 
energies just above the $\Omega$-production thresholds. At first, similar dynamic effects are 
observed for the $\Lambda$-hyperons as in the $\bar{p}$-induced reactions. They show the 
expected broad spectrum in rapidity (dashed curves) due to the enhanced multiple re-scattering. 
Latter causes also here their abundant capture inside the target nucleus, as shown by the 
dotted-dashed curves. The rapidity distributions of the cascade particles (dotted curves) 
are obviously peaked around the beam-value, but there is a significant contribution to 
lower rapidities too. This feature is again due to the secondary scattering, as discussed 
in previous works~\cite{Gaitanos14}. The production of bound cascade hyperons (dotted-dotted-dashed 
curves) is now enhanced. This effect induces the formation of exotic $\Xi$-hypernuclei (for 
more details on this task see Ref.~\citep{Gaitanos14}). 

%%%%%%%%%%%%%%%%%%
\begin{figure}[t]
\vspace{-2cm}
\begin{center}
\includegraphics[clip=true,width=1.0\columnwidth,angle=0.0]{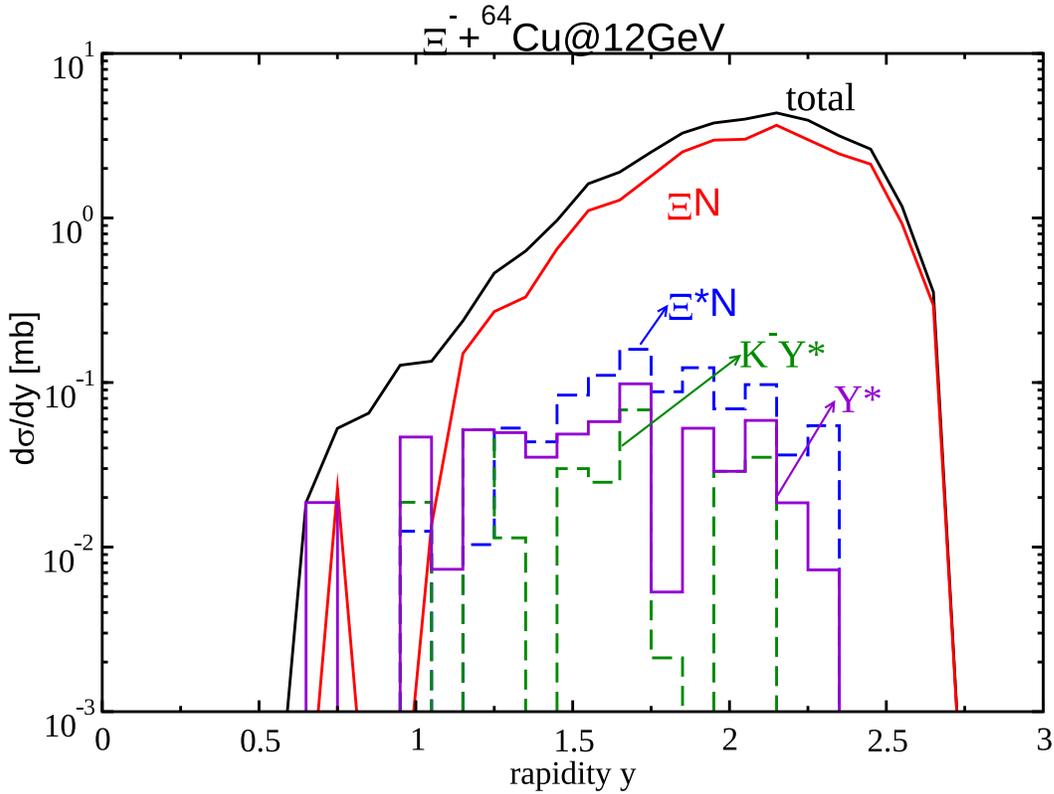}
\vspace{-1.0cm}
\caption{\label{fig6}
Rapidity distributions for $\Omega$-hyperons in the reaction $\Xi^{-}+{}^{64}Cu$ 
at $12$ \GeV~incident energy. Rapidity spectra for the total $\Omega$-yield 
(thick-solid black curve) and some particular contributions (thick-solid red: $\Xi N$, 
dashed blue: $\Xi^{\star}N$, thick-dashed green: $K^{\star-}Y^{\star}$, 
solid-violet: $Y^{\star}$) to the total spectrum are shown.
\vspace{-1.0cm}
}
\end{center}
\end{figure}
%%%%%%%%%%%%%%%%%%
The most interesting part for the present work is the thick-solid curves in Fig.~\ref{fig5}, 
which show the rapidity yields of the produced $\Omega$-baryons. The formation 
dynamics of the decuplet-particles here is similar to the dynamical production of the 
cascade particles in $\bar{p}$-induced reactions (for comparison see Fig.~\ref{fig4} again). 
However, the peak of the rapidity spectra of the $\Omega$ baryons is now located to much 
higher energies. In particular, the probability of bound $\Omega$-particles inside the residual 
nucleus is very low. These different dynamical formations between the $\Xi$- and $\Omega$-particles 
have physical reasons beyond the trivial ones (slightly different target masses and beam-energies). 
The decuplet particles are much heavier and carry one additional strangeness degree of freedom. 
Latter property causes multi-particle final states in many secondary processes of 
$\Omega$-production due to strangeness conservation. For instance, in binary collisions between 
the cascade-beam with other nucleons three final-state particles are required to conserve 
strangeness and baryon numbers. This leads to rather high threshold energies. The 
$\Omega$-production thresholds are also high in other secondary processes between 
the abundantly produced antikaons $K^{-}(S=-1)$ with hyperons or hyperonic resonances 
$\Lambda, \Sigma, Y^{\star}(S=-1)$. Thus, the $\Omega$-particles are produced with relatively 
high energies. The probability of secondary $\Omega$-scattering is low and they escape most 
likely the nucleus. 

Another interesting result is, that the $\Omega$-formation is pronounced largely in 
$\Xi$-nucleus collisions relative to the antiproton-induced reactions. In fact, 
the $\Omega$-production cross sections are in the range between $0.7-3.5$ \mb~for the 
incident $\Xi$-energies under consideration. This arises from the rather high cross section 
values of secondary scattering ranging in the \mb-region. This is manifested in 
Fig.~\ref{fig6}. It shows the contributions of various channels to the total $\Omega$-yield 
for the $\Xi^{-}+{}^{64}Cu$-reaction at an incident $\Xi$-energy of $12$ \GeV. The high momentum 
part of the produced $\Omega$-particles comes from primary collisions between the $\Xi$-beam 
and target nucleons. However, as one can see in Fig.~\ref{fig6}, the second-chance collisions 
involving antikaonic and hyperonic resonances contribute considerably to the $\Omega$-production 
over a broad longitudinal momentum, even for the intermediate mass number $A=64$ of the 
target nucleus. 

%%%%%%%%%%%%%%%%%%
\begin{figure}[t]
\vspace{-2cm}
\begin{center}
\includegraphics[clip=true,width=1.0\columnwidth,angle=0.0]{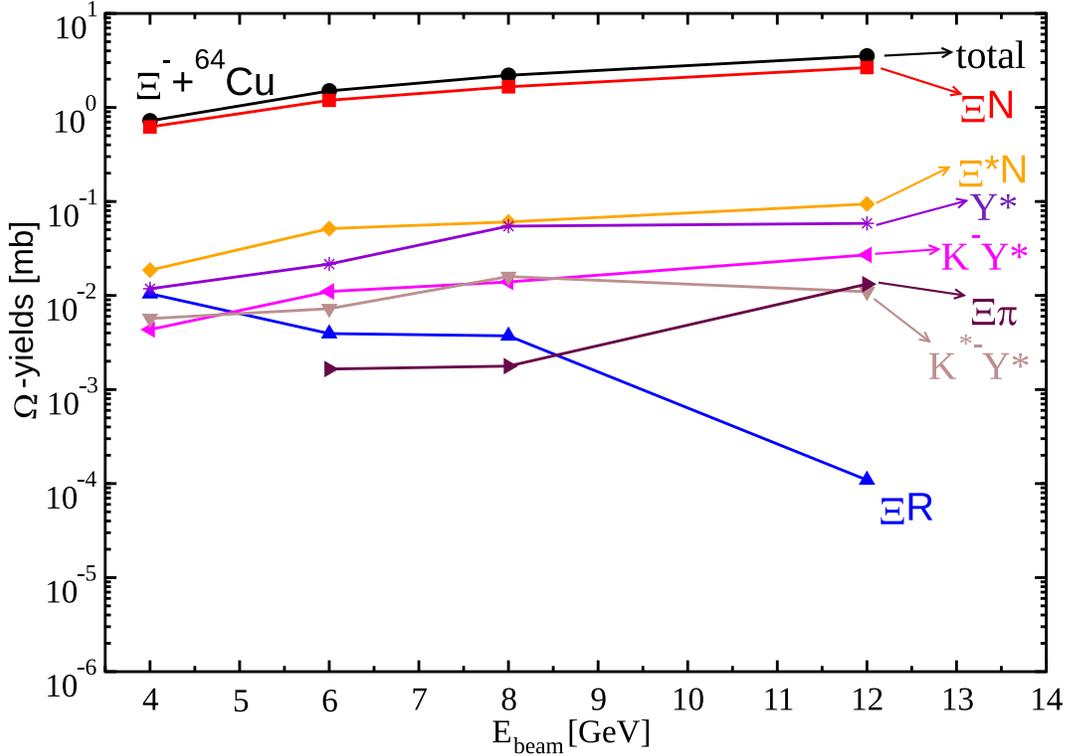}
\vspace{-1.0cm}
\caption{\label{fig7}
$\Omega$-production yields as function of the incident energy (in units of \GeV) 
for $\Xi^{-}+{}^{64}Cu$-reactions. (circles) total yield, 
(squares) $\Xi N$-contribution, (diamonds) $\Xi^{\star} N$-contribution, 
(stars) $Y^{\star}$-contribution, (left triangles) $K^{-}Y^{\star}$-contribution, 
(right triangles) $\Xi\pi$-contribution, 
(bottom triangles) $K^{\star -}Y^{\star}$-contribution, 
(top triangles) $\Xi R$-contribution. 
\vspace{-1.0cm}
}
\end{center}
\end{figure}
%%%%%%%%%%%%%%%%%%
A more complete picture is given in Fig.~\ref{fig7}, where the total $\Omega$-production yield 
and the most important contributions to it are displayed as function of the incident energy 
of the secondary cascade beam. The $\Xi$-nucleon binary collisions dominate over this 
energy range, as 
expected. The other secondary processes with the major contribution to the $\Omega$-multiplicity 
involve scattering between cascade resonances ($\Xi^{*}$) with nucleons and scattering between 
antikaons ($K^{-}, K^{\star -}$) with hyperonic resonances ($Y^{\star}$). Furthermore, with 
increasing beam energy channels with higher-mass resonances open, in particular, those 
channels with 
the hyperonic resonances. Therefore, the contributions from $\Xi R$-scattering 
(with $R$ being non-strange resonances) drops as the energy increases. Finally, the secondary 
scattering between cascade particles and non-strange mesons ($\Xi\pi$-channel) opens at higher 
energies only. In general, secondary processes with a hyperonic resonance $Y^{\star}$ 
in the initial channel together with the $\Xi N$-collisions mainly contribute to the 
production of $\Omega$-baryons. 
We predict total $\Omega$-production yields of several \mb~in the second-step 
$\Xi$-induced reactions and conclude the importance of the \panda-proposal towards the 
investigation of multi-strange in-medium hadronic interactions. 
In particular, as pointed out in the previous Sec.~\ref{sec2}, many theoretical 
models, for instance the chiral EFT calculations, 
provide us with elastic and quasi-elastic cross sections for scattering processes 
between the cascade baryon ($\Xi$) and the lighter $\Lambda$- and 
$\Sigma$-hyperons. Due to the importance of these secondary binary collisions to 
the dynamical formation of $\Omega$-baryons, the $\Xi$-nucleus reactions of the 
\panda-experiment may serve to better constrain the still existing uncertainties 
in these theoretical approaches. 

%%%%%%%%%%%%%%%%%%%%%%%%%%%%%%%%%%%%%%%%%%%%%%%%%%%%%%%%%%%%%%%%%%%%%%%%%%%%%%%%%%%
\section{\label{sec4} Summary and conclusions}
%%%%%%%%%%%%%%%%%%%%%%%%%%%%%%%%%%%%%%%%%%%%%%%%%%%%%%%%%%%%%%%%%%%%%%%%%%%%%%%%%%%

In summary, we have continued our previous investigations on multi-strangeness 
hypernuclear physics towards the decuplet-sector of SU(3) symmetry by considering 
the propagation and formation of the heavy $\Omega$-baryon in hadronic reactions 
relevant for FAIR. That is, antiproton-induced reactions supplemented by a 
secondary beam of $\Xi$-particles on an additional nuclear target. The theoretical 
description of these dynamical processes has been performed within the 
microscopic transport approach extended by the relevant $\Omega$-production channels, 
as far as phenomenological information was available. 

At first, we have studied the elementary processes leading to the formation of 
$\Omega$-hyperons. The primary channel consists of $N\bar{N}$-annihilation into 
$\Omega\bar{\Omega}$, for which theoretical estimations exist in the literature. 
The predicted $N\bar{N}\to\Omega\bar{\Omega}$-cross sections are too small 
with respect to other annihilation processes. In particular, the $\Omega$-production 
cross sections take values of a few \nb~only, which is approximately six orders 
of magnitude less than the nucleon-antinucleon annihilation into mesons. On the 
other hand, secondary re-scattering between antikaons and strangeness resonances 
occurs with much higher probability in the \mb-regime. 

In $\bar{p}$-nucleus reactions the formation probability of 
$\Omega$-particles is a very rare process. For this type of reactions we estimate 
$\Omega$-production yields of a few \nb~only. It turns out that secondary processes 
only do create $\Omega$-hyperons with a vanishing contribution from the primary 
$N\bar{N}\to\Omega\bar{\Omega}$. 

Our calculations, however, support the two-step reaction mechanism at \panda~for 
the observation 
of $\Omega$-hyperons. To be more precise, using the produced cascade-particles of the 
first-step $\bar{p}$-nucleus collisions as a secondary beam, additional transport 
calculations for the realization of $\Xi$-induced reactions were performed. A high 
energy $\Xi$-beam was utilized to overcome the high $\Omega$-production thresholds. 
At first, an abundant production of the lighter $\Lambda$- and $\Xi$-hyperons was 
observed in consistency with previous studies. The most important result was the 
formation of the $\Omega$-particles in $\Xi$-nucleus collisions with an observable 
probability. It is still an open question how the $\Omega^{-}$-hyperon can be 
observed experimentally probably by requiring reconstruction from coincidence 
experiments and particle correlations. However, the production cross sections in 
the \mb-region, 
as estimated from the present analysis, indicate a high production rate. The dynamical 
formation of these heavy hyperons was investigated too. For multiple $\Omega$-formation 
the $\Xi N$-processes gives the highest contribution, as expected for these reactions 
with the $\Xi$-beam. As an interesting feature the contributions from other secondary 
channels was significant even with a intermediate-mass target nucleus. It would be a 
challenge to explore such dynamical hadron-nucleus reactions experimentally too. 
It will be important to constraint better the physical picture of multi-strangeness 
elementary processes. 

We conclude with pointing out the great opportunity of the future activities at FAIR to 
understand deeper the 
still little known high strangeness sector of the hadronic equation of state. Note that 
the strangeness sector of the baryonic equation of state is crucial for our knowledge 
in nuclear and hadron physics and astrophysics. 
For instance, hyperons in nuclei do not experience Pauli blocking within the Fermi-sea of 
nucleons. Thus they are well suited for explorations of single-particle dynamics. 
In highly compressed matter in neutron stars the formation of particles 
with strangeness degree of freedom is energetically allowed. Of particular 
interest are hereby the $\Lambda$-,$\Sigma$-, $\Xi$- and $\Omega$-hyperons 
with strangeness S=-1,-2 and -3, respectively. As shown in recent 
studies~\cite{NSEoSHyp4,Weissenborn:2011kb}, 
these hyperons modify the stiffness of the baryonic EoS at high densities considerably leading 
to the puzzling disagreement with recent observations of neutron stars in the range of 2 solar 
masses.

\section*{Acknowledgement}
Supported in part by DFG, contract Le439/9, BMBF, contract 05P12RGFTE, GSI Darmstadt, and 
Helmholtz International Center for FAIR.

\section*{References}

%\begin{thebibliography}{10}

\end{document}